\documentclass[conference]{IEEEtran}
\IEEEoverridecommandlockouts
\usepackage{cite}
\usepackage{amsmath,amssymb,amsfonts}
\usepackage{algorithmic}
\usepackage{graphicx}
\usepackage{soul} % \hl{}
\usepackage{textcomp}
\usepackage{xcolor}
\usepackage[inline]{enumitem}
\usepackage{subcaption}
\usepackage{calc}
\usepackage{float}
\usepackage{calc}
\usepackage{hyperref} % Sean added this package to make the links show up

\usepackage[disable]{todonotes}
\usepackage{listings}
\lstdefinestyle{mystyle}{
    commentstyle=\color{gray}\ttfamily,
    numberstyle=\tiny\color{darkgray},
    stringstyle=\color{purple},
    basicstyle=\footnotesize,
    breakatwhitespace=false,         
    breaklines=true,                 
    captionpos=b,                    
    keepspaces=true,                 
    numbers=left,                    
    numbersep=5pt,                  
    showspaces=false,                
    showstringspaces=false,
    showtabs=false,                  
    tabsize=1,
    upquote=true,
    frame=single,
    morekeywords={as, include, printf, a}
}
\newcommand{\eg}{\emph{e.g.},}

\newcommand{\ie}{\emph{i.e.},}

\newcommand{\cf}[1]{\emph{cf.} Sec. \ref{#1}}
\newcommand{\ourvspace}{\vspace{-0.4em}}

\newcommand{\ourvspacefig}{\vspace{-1.5em}}
\newcommand{\aftercaptionvspace}{\vspace{-1.0em}}

\newcommand{\codefont}[1]{\texttt{#1}}

\def\BibTeX{{\rm B\kern-.05em{\sc i\kern-.025em b}\kern-.08em
    T\kern-.1667em\lower.7ex\hbox{E}\kern-.125emX}}

\begin{document}

\bstctlcite{IEEEexample:BSTcontrol}

\title{
% Not sure if this is the final title. Still thinking.
%Towards Lightweight Multi-workflow Data Integration using Provenance and Data Observability
Towards Lightweight Data Integration using Multi-workflow Provenance and Data Observability
%: A Deep Learning Use Case

\thanks{
Preprint of a paper accepted at the 19th IEEE International Conference on e-Science 2023. Please cite it as follows:
R. Souza, T. Skluzacek, S. Wilkinson, M. Ziatdinov, R. da Silva, ''Towards Lightweight Data Integration using Multi-workflow Provenance and Data Observability,`` Accepted at the 19th International Conference on eScience (eScience), Limassol, Cyprus, 2023. \\
Correspondence: Renan Souza - contact@renansouza.org
}
\ourvspace{}
}

\author{
\IEEEauthorblockN{
    Renan Souza,
    Tyler J. Skluzacek,
    Sean R.     Wilkinson,
    Maxim Ziatdinov,
    Rafael Ferreira da Silva}
\IEEEauthorblockA{
    National Center for Computational Sciences, Oak Ridge National Lab, Oak Ridge, TN, USA\\
    \{souzar, skluzacektj, wilkinsonsr, ziatdinovma, silvarf\}@ornl.gov}
    \vspace{-2.5em}
}

\maketitle

\begin{abstract}
Modern large-scale scientific discovery requires multidisciplinary collaboration across diverse computing facilities, including High Performance Computing (HPC) machines and the Edge-to-Cloud continuum. 
%One example of this collaboration involves chemists who capture massive volumes of data using an electron microscope at the Edge. They then utilize distributed deep learning techniques on an HPC machine. The resulting models are validated by experts in the cloud, contributing to fine-tuning the microscope in a continuous feedback loop. 
Integrated data analysis plays a crucial role in scientific discovery, especially in the current AI era, by enabling Responsible AI development, 
%(\eg{} explainability, bias assessment, model impact analysis)
FAIR, Reproducibility, and User Steering. However, the heterogeneous nature of science poses challenges such as dealing with multiple supporting tools, cross-facility environments, and efficient HPC execution.
Building on data observability, adapter system design, and provenance, we propose MIDA: an approach for lightweight runtime \underline{M}ulti-workflow \underline{I}ntegrated \underline{D}ata \underline{A}nalysis. 
MIDA defines data observability strategies and adaptability methods for various parallel systems and machine learning tools. 
With observability, it intercepts the dataflows in the background without requiring instrumentation while integrating domain, provenance, and telemetry data at runtime into a unified database ready for user steering queries.
We conduct experiments showing end-to-end multi-workflow analysis integrating data from Dask and MLFlow in a real distributed deep learning use case for materials science that runs on multiple environments with up to 276 GPUs in parallel.  We show near-zero overhead running up to 100,000 tasks on 1,680 CPU cores on the Summit supercomputer.
\end{abstract}

%\ourvspace{}
\section{Introduction}

Modern science is characterized by its multidisciplinary and interconnected nature. This has become evident during the COVID-19 pandemic when numerous teams collaborated across heterogeneous computing facilities, such as High Performance Computing (HPC) environments and the Edge-to-Cloud continuum. 
These teams employ diverse tools for unified data processing and analysis in a science campaign.
% These teams utilize various tools for data processing and analysis toward a unified goal in a science campaign. 

Materials scientists, for instance, are leveraging the power of electron microscopes to understand materials' structure on the atomic level to guide the design of new functional materials for applications in biotechnology, advanced electronics, and energy storage. 
For example, they can better understand battery materials at the atomic level, providing critical insights about their functioning, and thus guiding the design of long-lasting energy storage solutions. 
These experiments involve collecting large volumes of image data, often requiring immediate processing at the Edge to guide ongoing experiments. 

It is already commonplace to employ Deep Learning (DL) in science~\cite{souza2020workflow}. Typically trained on HPC machines, these models can be independently deployed or integrated into an active learning pipeline to steer scientific discovery~\cite{ziatdinov2022atomai}. 
Similar scenarios are progressively pervasive in various domains, such as Oil \& Gas, High-energy Physics, and Healthcare~\cite{gil_intelligent_2018,zurawski2022high,rosenthal2022building}.

\begin{figure*}[!t]
    \centering
    \includegraphics[width=0.85\linewidth]{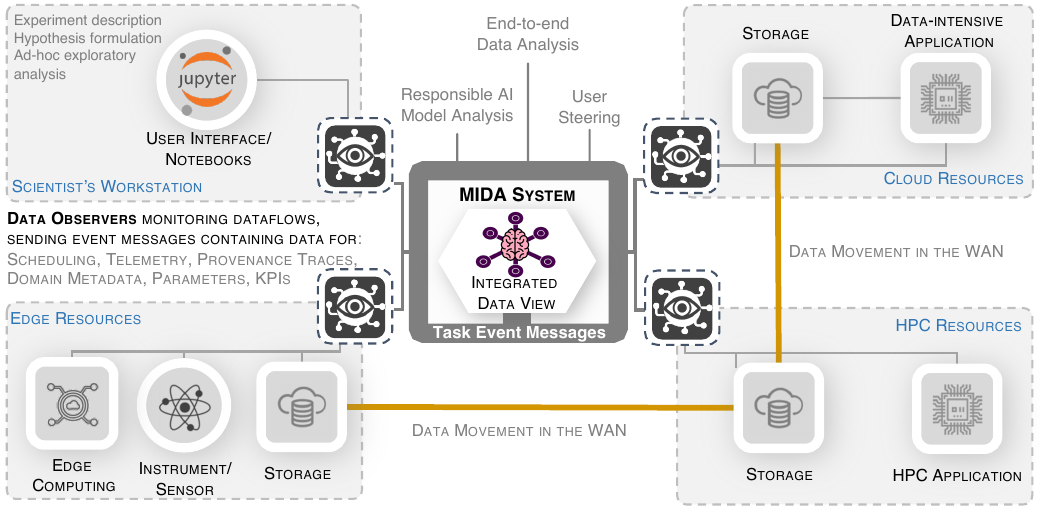}
    % \ourvspace{}    
    \caption{%
        Solution high-level overview, with data observer agents monitoring the dataflows in each environment in use.
    }    
    \label{fig:overview}
    \aftercaptionvspace{}  
\end{figure*}

%\ourvspace{}
\subsection{Motivation}

In the current AI era, there is a growing need for responsible AI model development, particularly for scientific discovery. Integrated data analyses are necessary for critical capabilities, including transparency, explainability, bias assessment, FAIR (Findability, Accessibility, Interoperability, and Reusability), reproducibility, and user steering (\eg{} monitor, analyze, or fine-tune the experiments at runtime)~\cite{souza_keeping_2019,souza2020workflow}.
Provenance (a.k.a. lineage) data management provides valuable support for reproducing, tracing, understanding, and explaining data, models, and their transformation processes~\cite{liao2023ai,mattoso_scientific_2010}.
% Provenance (a.k.a. lineage) plays a key role in managing data, models, and their transformation processes, thus supporting integrated data analyses. It provides valuable support for large-scale computational scientific experiments by facilitating reproducibility, traceability, and result data understanding~\cite{mattoso_scientific_2010,ccpe_first_prov_challenge_2008, buneman2019data}.

The significance of provenance becomes evident when considering concrete analysis scenarios such as: 
\textit{``describe the whole training process, with its parameters, model architecture, used datasets, and data transformations of the 10 models with lowest loss"}, or \textit{``show the dataset metadata and parameters of the microscope that generated the raw dataset used in the training processes that ended up in GPU memory overflow"}.
These analyses aid experiment understanding and provide insights at runtime that enable users to adapt and guide their experiments towards successful outcomes~\cite{souza_keeping_2019}.

Nevertheless, the heterogeneous nature of modern science poses challenges~\cite{wcs2022,dare19,sigmodpanel_2020}. First, the issue of cross-facility execution environments arises when scientific campaigns traverse multiple sites with diverse computing facilities. 
%There is a need for a  unified data layer that seamlessly connects these components, overcoming the hurdles of the environments' heterogeneity.
Second, scientists and engineers use a multitude of tools for data processing and analysis, such as Parallel Computing Systems (PCSs) or scripts with various supporting tools; thus, making data integration hard. 
Last, enabling integrated data analysis in HPC is complex, as solutions cannot incur high overhead. 
% Addressing these challenges is critical in driving scientific discovery~\cite{wcs2022,sigmodpanel_2020}.

%\ourvspace{}
\subsection{Problem, Objective, and Contributions}

The computationally processable parts of a scientific campaign are considered \textit{workflows}, whether managed by a PCS or not. The workflows consist of interconnected activities, forming a \textit{multi-workflow}~\cite{provlake_escience_2019}.
Despite the individual software and hardware requirements, independent execution control, and execution on various environments for each workflow, the data must be analyzed in a unified and integrated way.
Also, the integration must occur at runtime to accelerate \textit{time-to-insight}, as users cannot wait until the end of long-running workflows before initiating the integration and analysis process.

In summary, the problem addressed in this work is:
% \noindent
% \textbf{\underline{Problem}}: 
\textit{How to enable runtime integrated data analysis in multi-workflows aware of the wide variety of use cases, each potentially using different PCSs or various tools in scripts?
}

% \noindent \underline{\textbf{Solution}}: 
This paper introduces MIDA (\underline{\textbf{M}}ulti-workflow \underline{\textbf{I}}ntegrated \underline{\textbf{D}}ata \underline{\textbf{A}}nalysis), an approach for \textit{\textbf{runtime data integration}} that builds upon data observability~\cite{observability_sigmod2021} and provenance~\cite{mattoso_scientific_2010}. 
Provenance-based data views have proven valuable both in single~\cite{dfanalyzer_2020,souza_keeping_2019} and multi-workflows~\cite{souza2020workflow, provlake_escience_2019}.
MIDA creates an integrated data view at runtime, seamlessly connecting the data processed by multi-workflows. The view is lightweight, 
% by utilizing workflow provenance based on W3C PROV~\cite{W3CPROV} as its foundational framework. 
encompassing parameters, meaningful domain metadata, KPIs (\eg{} ML performance, figures of merit), task and telemetry data, and pointers to the heavy data files in federated storages, all interconnected via W3C PROV-based~\cite{W3CPROV} provenance.

MIDA goes beyond previous works with a novel approach that establishes data observability strategies to monitor dataflows in the background and build a comprehensive integrated data view, as depicted in Figure~\ref{fig:overview}.
The view is built without requiring instrumentation and is materialized in a database. It enables users to run steering queries to inspect 
a global and detailed view of the workflows and their data at runtime.
 Another key aspect is that by following an adapter design, MIDA defines adapter methods for the integration of data coming from a multitude of tools that support scientific discovery, regardless if they already manage provenance data. 

Therefore, the objective is 
\textit{to accelerate time-to-insight via runtime integrated data analyses in multi-workflow  embracing various domains and use cases that rely on: (i)~one or many PCSs; (ii) scripts with one or many data capturing tools; or (iii) any combination of (i) and (ii).}
% Adapters for existing systems and an API to develop more adapters are available.}

Specifically, this work makes the following contributions:

% \noindent
% \underline{\textbf{Contributions}}:

\begin{enumerate}[leftmargin=1em]
%\begin{enumerate}[label=(\arabic*),nolistsep,wide]

\item 
System architecture and design principles for lightweight runtime data integration of diverse scientific discovery tools.
% System architecture and design principles for lightweight integration of data  from a multitude of tools for scientific discovery.

% \item Pioneer approach for enabling runtime capture and analysis of provenance data for Dask workflows.
\item A generic method that enables runtime provenance management in PCSs that lack native support, as an initial step towards integrating with MIDA. This approach is demonstrated in Dask, a widely used PCS, making this the pioneering work to enable provenance for Dask, regardless of whether it involves single- or multi-workflows.

\item Lessons learned from large-scale experiments on Summit and Crusher supercomputers, involving integrated data analyses in scientific deep learning using up to 276 GPUs and over 100,000 parallel tasks across 1,680 CPU cores.

\end{enumerate}

%\ourvspace{}
\section{Background} 
\label{sec:background}

\subsection{Challenges in Large-scale Scientific Data Analysis}
\label{sec:challenges}

\textbf{Cross-facility.} Scientific campaigns often involve multiple sites with diverse computing facilities, leading to distributed, federated, and heterogeneous execution environments~\cite{cross_facility_wf_2021,wcs2022}. These environments  include HPC machines for distributed DL, cloud clusters for data preprocessing, analysis, and visualization, and Edge devices for data processing close to instruments or sensors. A major challenge is to establish a unified data layer that interconnects cross-facility components within a campaign and enables efficient data integration.

\textbf{HPC.}
HPC environments, including leadership supercomputers~\cite{top500}, are crucial for executing demanding workloads such as distributed DL or simulations. 
As users cannot wait until the end of long-running workflows,
a challenge lies in enabling integrated data analysis at runtime without incurring significant performance overhead. The goal is to devise approaches that optimize data analyses while maintaining high scalability and ensuring efficient utilization of HPC resources.

\textbf{Multitude of Tools.}
Some users rely on a range of PCSs to run their experiments, including workflow management systems  (\eg{} FireWorks~\cite{jain2015fireworks}, Swift/T~\cite{wozniak2013swift}), parallel scripting libraries (\eg{} Parsl~\cite{babuji2019parsl}, Dask~\cite{dask}), and big data frameworks (\eg{}  Spark~\cite{spark}).
These systems handle parallel task scheduling and enable monitoring, managing data for scheduling and optionally for resource usage and provenance. Other users write their parallel code without a PCS and still require monitoring and provenance. They instrument their scripts with tools for profiling, telemetry data collection, logging, provenance, or dataflow analysis tools. In machine learning, users rely on specific tools, like MLFlow~\cite{zaharia_accelerating_2018}, for tracking training progress.
 However, given the diverse range of PCSs and runtime data capture tools, adopting a one-size-fits-all approach is impractical. 
 Thus, the challenge lies in designing a system capable of embracing a wide variety of tools while addressing the needs of different domains and use cases.

\subsection{Multi-workflows and Provenance}

% \noindent
% \textbf{Scientific Method and Cross-facility.}
% In the scientific method, scientists and engineers continuously formulate questions and hypotheses, conduct experiments, analyze data, and report conclusions\cite{mattoso_scientific_2010}. As modern science progressively becomes multidisciplinary and requires more computing with diverse requirements, both in terms of cross-facility execution environments and a variety of supporting tools.

% \noindent
% \textbf{Single- and Multi-workflows.}
Applications of interest in this work include parallel computational simulations, big data analytics, and distributed ML training. To ease the complexity of running these applications, a common practice is to model them as a \textit{single-workflow}, or workflow for short. A workflow consists of interconnected \textit{activities} that process data. Activities can be either a black-box program, program functions, methods, or code blocks within a script. Workflows can be executed using a PCS or as scripts (or Notebooks). In a given campaign, users may utilize a combination of PCSs and scripts to run various workflows across different facilities, with each execution having its own autonomous control. A \textit{multi-workflow} is a composition of these single-workflows, where the data generated by one workflow are consumed by another. The analysis of \textit{multi-workflow data} necessitates an integrated approach~\cite{provlake_escience_2019}.

Workflow provenance traces can connect heterogeneous components in a science campaign and establish integrated data analysis~\cite{mattoso_scientific_2010,provlake_escience_2019}. Leveraging W3C PROV core classes (\codefont{Activity} and \codefont{Entity}) and their relationships (\codefont{used} and \codefont{wasGeneratedBy})~\cite{W3CPROV}, we create links within single- or multi-workflows. 
%For instance, an ML workflow with \codefont{activity1} (model training on \codefont{entity1} to generate \codefont{entity2}) and \codefont{activity2} (evaluation of \codefont{entity2} to produce evaluation metrics \codefont{entity3}) can be chained as \codefont{entity3 wasGeneratedBy activity2 used entity2 wasGeneratedBy activity1 used entity1}.
%
%Extending this approach, a multi-workflow provenance trace is formed when \codefont{entity1}, the output of a preceding data preparation workflow, is used as the first input in the training workflow \cite{provlake_escience_2019}. 
%This comprehensive trace documents the entire ML lifecycle from raw data acquisition to trained models and their inference. 
Additional semantic richness is achieved using other PROV elements %(to represent, \eg{} a user who trained a model)
and domain-specific extensions~\cite{souza2020workflow}.
%Implementation-wise, workflow management systems or data capture systems utilize DBMSs \todo{did I define DBMS?} with various data models such as relational, document-oriented, graph-oriented, or triple stores to manage the provenance data~\cite{jain2015fireworks, dfanalyzer_2020, provlake_escience_2019, prov_io}.

%\ourvspace{}
\subsection{Data Managed in PCSs and Runtime Data Capture Tools}
\label{sec:data_pcs_tools}

PCSs aim at facilitating the management of parallel applications. They handle data for scheduling, such as task metadata (\eg{} task states, parameters, output logs, and resource information).
% like the node in a cluster where the task executes). 
Some PCSs also support performance monitoring, providing insights into resource usage and elapsed times per task. Workflow management systems, in particular, typically extend their capabilities to include provenance data analysis, tracking the data used and generated by each task, and forming workflow provenance traces. Combining data for scheduling, performance monitoring, domain, and provenance has proven effective in enabling users to gain a comprehensive understanding of their large-scale computational experiments~\cite{souza2020workflow, souza_keeping_2019, dfanalyzer_2020}. For cases where a PCS is not employed but a workflow still processes data that requires analysis, runtime data capture tools \cite{dfanalyzer_2020,provlake_escience_2019} offer similar data analysis and monitoring as PCSs but do not manage parallel tasks directly.

%\ourvspace{}
\subsection{Data Observability}

Data observability is emerging as a paramount component for monitoring and maintaining large-scale distributed systems, ensuring QoS~\cite{observability_sigmod2021}. It surpasses basic monitoring capabilities by collecting telemetry and usage information at runtime, allowing for deeper insights into system correctness and performance. This data-intensive approach needs user involvement in monitoring, inspecting disruptive events, and making runtime decisions. The management of data observability encompasses four main categories: metrics, events, logs, and execution traces~\cite{observability_sigmod2021}. By integrating data observability with domain-specific and provenance data, its scope can expand beyond system health monitoring. This integration involves incorporating domain-specific KPIs, figures of merit, and multi-workflow provenance traces that capture the structured dataflow within multi-workflows. Combining data observability and provenance data techniques has the potential to provide valuable context and enhance the understanding of results and performance within multi-workflows.

%\ourvspace{}
\section{Integrated Data Analysis using Multi-workflow Provenance and  Observability}
\label{sec:mida}

\begin{figure}[!t]
    \centering
    \includegraphics[width=1.0\linewidth]{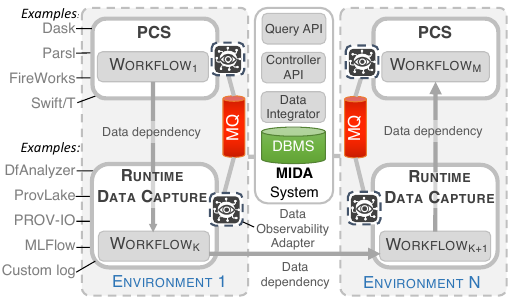}
    \ourvspacefig{}    
    \caption{%
        System Architecture for MIDA, with various Parallel Computing Systems and Runtime Data Capture tools. 
    }
    %TODO: Swift/T?  need to double check its logging mechanisms
    \label{fig:architecture}
    \aftercaptionvspace{}  
\end{figure}

To tackle the challenges in this work, we begin by proposing a system architecture (Fig. \ref{fig:architecture}) and design principles for MIDA.

To address the cross-facility challenge, the architecture relies on distributed, portable, and loosely-coupled data observers that plug into PCSs or runtime data capture tools running  across the execution environments where the workflows execute. 
They are loosely coupled because they are designed not to interfere with the critical execution path of a PCS or runtime data capture tool. They observe the dataflows as a background service and send events to a message queue (MQ) system.
%As a background service, they can start or stop observing based on user commands.
They are portable and can be deployed on heterogeneous resources ranging from scientists' desktops to large HPC machines using cross-platform technologies.

To ensure efficient execution in HPC environments, a key aspect is the decoupled design of the system. By minimizing interference with task scheduling, the performance overhead is significantly reduced in most scenarios. Even in cases where interference may occur, the overhead remains small due to the observers' limited processing capabilities, focused on monitoring dataflows and filtering relevant data within a stream. A Database Management System (DBMS) manages the integrated data view, but the data observers do not access it directly. Instead, heavier data integration operations and DBMS accesses are isolated from the main computation. Observers only send asynchronous event messages to the MQ. Employing a technique from prior work~\cite{souza2020workflow}, multiple messages are grouped in buffers and sent in bulk, effectively reducing communication overhead. For performance-sensitive workflows, the recommended approach is to deploy an MQ system within the workflow's environment, enabling local communication and further minimizing overhead.

To address the challenge of supporting a multitude of tools, 
MIDA follows an adapter system design~\cite{freeman2004head}. 
We define adapter methods for observability and integration of data from various data capture tools and PCSs, as discussed in Sec.~\ref{section:main_arch}. 

%\ourvspace{}
\subsection{Task Metadata}
\label{section:task_metadata}

% Task-oriented programming models have been extensively used by PCSs for years. %~\cite{raicu2008many}.
A task represents the concrete execution of a workflow activity.
%Depending on the implementation, it can represent a fit of an ML model or an iteration in a time-step simulation or the execution of a scientific program that uses certain arguments and generates a set of files.
Tasks are the finest element managed by PCSs (\eg{} to schedule them in a cluster). 
Runtime data capture tools also use tasks 
%(or related terms, such as \textit{data transformation execution} \cite{dfanalyzer_2020, provlake_escience_2019}) 
as the main concept in their data schema. 
The integrated data view is essentially composed of sets of task metadata with their used task parameters or datasets and generated results or datasets. Standard output logs and errors, statuses, and timestamps also compose task metadata~\cite{COLEMAN202216}. 
In terms of W3C PROV concepts \cite{W3CPROV}, a task is an \codefont{Activity} that links with 
\codefont{Entities}, \ie{} parameters, results, or datasets, via the relationships \codefont{used} and \codefont{wasGeneratedBy}.

Since proposing yet another workflow provenance schema is not a goal in this paper, we reuse a provenance schema that can establish multi-workflow provenance traces~\cite{provlake_escience_2019}. This allows tracking data dependencies within a single-workflow and establishing connections between entities generated or used between workflows. Besides, each task in the data view includes relevant details such as system information, execution environment, user, science campaign (which groups workflows in a multi-workflow scenario), and resource utilization.
% The implementing data observability adapters need to provide the data to populate this information for each of their tasks.
% A concrete example is provided in the next section, with implementation details.

%\ourvspace{}
\subsection{Data Observers}
\label{section:main_arch}

One solution to support the multitude of tools could be to force all tools to follow the same API standard or to generate data following the same data schema. Nevertheless, given the high heterogeneity of these tools and their fast development and adoption by various use cases, it is nearly impossible to establish a one-size-fits-all standard. Rather, our solution relies on adapters for each tool or PCS.
Data observers are background services that capture metadata by observing dataflows during workflow execution. 
They implement the adapter between the process responsible for observing task status changes and the tool or system being observed.  The data providers must meet the following requirements:

\textit{Provide access to the basic task metadata}. 
The essential metadata required include the used and generated data by the tasks, encompassing parameters, results, and datasets, as they serve as the foundation of the data view and enable multi-workflow provenance. Timestamps and statuses have more flexible requirements and can be generated by the observer.

\textit{Provide awareness of task state change}.
Observers need to detect changes in task states (\eg{} from submitted to running then finished), that affect runtime runtime monitoring. When a task state changes, the observer analyzes its metadata to determine if an interception is necessary, such as when the metadata contains relevant KPIs to monitor. If an interception is needed, the observer generates a task state change event message with the relevant metadata, captures telemetry data if enabled, and sends the message to an MQ system.

When a task undergoes a state change, there are two strategies for the observer to perceive it: 

\textit{External Polling Strategy.} 
The external polling strategy involves regularly polling an external task data component to detect task state changes. Components such as DBMSs, MQ systems, log files or API-based services can be polled to retrieve task metadata. This strategy applies to both PCSs and runtime data capture tools. Examples of PCSs include FireWorks~\cite{jain2015fireworks}, which utilizes a document DBMS, and Parsl~\cite{babuji2019parsl}, which logs task state changes in files. Runtime data capture tools like ProvLake~\cite{provlake_escience_2019} and PROV-IO~\cite{prov_io} use RDF stores. This strategy is preferred when the polled component is not used for scheduling. However, if the observer continuously polls a component used for scheduling, it may add overhead by competing with the scheduler, requiring  evaluation.

% \noindent
% \hangindent=0.7em 
\textit{Intra-PCS Scheduler Strategy.} 
When observers are situated within the internals of a task scheduler, which manages task metadata internally or lacks direct access to its task data management component, there can be significant overhead due to observability activities that may compete with scheduling. To address this, three approaches can be taken: (i)~the data provider can offer extensible plugins to customize task state changes within the scheduler; (ii)~a wrapper can be added on top of the scheduler to notify observers of task changes without modifying the PCS's code directly; (iii)~the PCS's code can be modified to generate notifications. The first approach is the most elegant, allowing PCS developers to identify extensible points in the scheduler. The third approach is often the most complex, involving modifications to a complex codebase and undesired coupling between the PCS and the observer. The second approach strikes a balance between the two, serving as a viable alternative when the PCS code is not extensible.

%\ourvspace{}
\subsection{Interfaces and Data Integrator}
\label{sec:data_observer_interface}

\noindent
\textbf{Data Observer Adapter Interface.} When creating a new data observer, two main methods should be implemented: \codefont{observe} and \codefont{callback}. The \codefont{observe} method follows a polling strategy, querying an external component associated with a data provider. For instance, in an MLFlow deployment using a relational DBMS, an adapter can employ SQL-based polling. The \codefont{callback} method determines the actions to take when a task state change is detected. It decides if the change has relevant data for the data integration. If the change is relevant, the observer builds a message and invokes a function responsible for transmission (\eg{} buffering and communication with the MQ). When using intra-PCS scheduler, only the \codefont{callback} method needs to be implemented, as the observability will depend on the PCS's scheduler. For example, Dask provides extensible plugins for this purpose. Further adapter implementation details are in Section \ref{sec:impl}.
%The main methods to implement when developing a new data observability adapter for a PCS or runtime data capture tool are: \codefont{observe} and \codefont{callback}. 
% The \codefont{observe} method implements the polling strategy to an external component for a given data provider, to implement the reaction whenever a task state change is identified. As a simple example, in an MLFlow deployment that uses a relational DBMS, a data observability adapter can employ a SQL-based polling service that keeps querying the database at every time unit looking for a change. 

% The \codefont{callback} method implements the logic that decides what to do when a task state change is identified. It can, for instance, decide if that task event change contains data to compose the integrated data view.
% If it is an interesting change, the observer builds a dictionary data structure with the message to be sent and calls a function that receives this dictionary and implements the message-sending mechanisms (\eg{} buffering and communication with the MQ system).
% In scenarios where the only observability strategy possible is to attach a PCS's scheduler internals, only the \codefont{callback}  method in the interface must be implemented, as the observability will take place differently, depending on the PCS's scheduler. For example, in Dask, developers can use its extensible plugins.

\noindent
\textbf{Controller API.}
This API consists of an initialization method and start/stop methods. During initialization, users register PCSs and runtime data capture tools in the campaign, providing necessary configuration variables, like file paths for logging or routes to a DBMS for polling. The start method launches the observers in the environment, while the stop method notifies observers to complete any ongoing operations (\eg{} closing DBMS connections) and gracefully terminate.
% A simple API with an initialization method and start and stop methods. In the initialization method, the user registers, before execution, the PCSs and the runtime data capture tools in the campaign, providing whatever configuration variables that apply. For example, if one uses a simple logging library to log data to a file, this file path needs to be provided. Similarly, if a DBMS will be polled by a data observer, routes to it need to be provided. Then, a start method will start up the observers in the environment and the stop will notify the observers so that they can finish any remaining observability operations (\eg{} close a DBMS connection or close a file being polled, flush any buffer) and gracefully stop.

\noindent
\textbf{Data Integrator.}
 While data observers are lightweight, performing limited processing, monitoring dataflows, and selecting relevant data, the data integrator generates multi-workflow traces and stores them in a persistent DBMS. It queries the DBMS to establish relationships between tasks in different workflow executions, ingesting task metadata as needed. This separation of concerns is important for overhead management. By combining asynchronous emissions from the data observers with buffering messages at the observers and data integrator, bulk insertions (both into the DBMS and the MQ), and indexing, overhead can be reduced.
 A Query API is provided to facilitate users and applications to query the DBMS.  It is available both in Python and as a RESTful HTTP API.
It allows OLAP queries (projections, filters, aggregations, and sorts) to the tasks' metadata. Listing \ref{listing:flowcept_query} has an example.
% This query call is translated to a native query of the DBMS used.
% More complex queries are currently available by querying the DBMS directly.
% This is the component that receives the event messages sent by the distributed data observers. We follow lessons learned in previous work \cite{provlake_escience_2019} to keep a clear separation of concerns. 
% While the data observers do very limited processing, only monitoring the dataflows and selecting relevant data to intercept,
% the data integrator is responsible for generating the multi-workflow traces and storing in a persistent DBMS. 
% Given a set of tasks in a certain workflow execution, the data integrator may query the DBMS looking for data of tasks in other workflow executions to enable the relationships between multiple workflow executions. Then the data integrator inserts new task metadata or updates existing ones (\eg{} when a task finishes, it updates with resulting data, statuses, and logs).
% Since the data observers are closer to the workflows being executed, the separation of concerns is important to isolate the observers from heavier data integration operations, hence reducing performance overhead.
% The data in the DBMS can be queried via a Query API, as we show in the validation section.

\noindent
\begin{minipage}{0.47\textwidth}
\vspace{0.35em}
\begin{lstlisting}[
    language=python,
    caption={ Simple Query API call to the DBMS. },
    label={listing:flowcept_query},
    keywords={projection, filter, sort, aggregation, limit}
]
result_set = query(
 projection=['used.layers','generated.loss'],
 aggregation=[('min','generated.loss')]
 filter={'campaign_id': 'my_campaign_1'},
 sort=[('ended_at', 'descending')],
 limit=100
)
\end{lstlisting}  
\ourvspace{}
\end{minipage}
\ourvspace{} 

\noindent

%\ourvspace{}
\section{Experimental Validation}

\begin{figure*}[!h]
    \centering
    \includegraphics[width=1.0\linewidth]{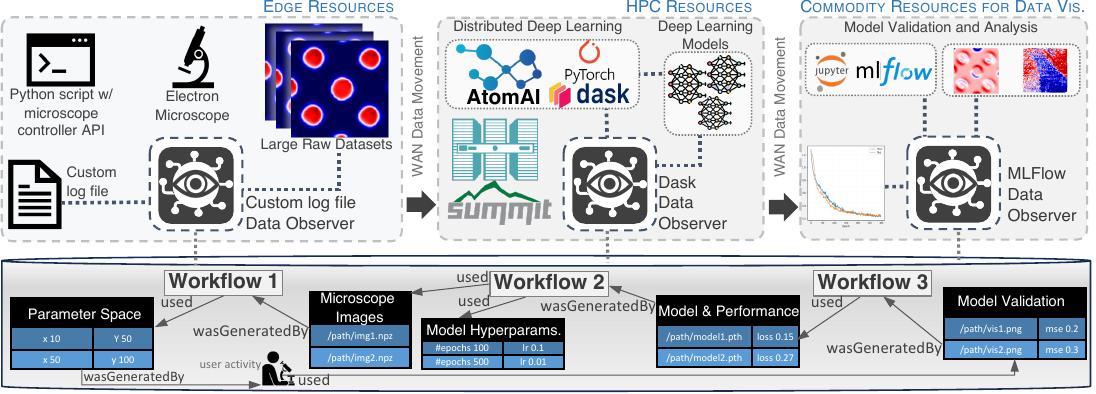}
    \ourvspacefig{}    
    \caption{%
        Electron microscopy use case, where the scientists tune the next position of the microscope based on results generated by a Deep Learning model, showing a human-in-the-loop aspect.  Multi-workflow provenance traces are captured by Dask, MLFlow, and custom log data observers and stored in an integrated data view materialized in a database.
    }
    %todo: y50 in parameter space shouldn't be capitalized
    \label{fig:usecase}
    \aftercaptionvspace{}  
\end{figure*}

%\ourvspace{}
\subsection{Motivating Use Case: Deep Learning on Microscopy}

In this work, we focus on electron microscopy deep learning and its application in various fields such as condensed matter physics, biology, and materials science. By utilizing AI models like deep convolutional neural networks (CNNs), researchers can leverage the power of CNNs to efficiently identify atomic species and defects, and track their evolution in materials. To support this, we employ AtomAI~\cite{ziatdinov2022atomai}, an open-source tool that simplifies the use of CNNs for tasks such as semantic segmentation of atomic images, generating descriptors from spectroscopy data, and predicting atomic species and positions.

Figure \ref{fig:usecase} illustrates a multi-workflow scenario with three distinct workflows running on different facilities and resources. In \textit{Workflow 1}, a Python script controls an electron microscope at the 
Center for Nanophase Materials Sciences at Oak Ridge National Laboratory (ORNL), generating large volumes of raw imagery data in the Edge. These data are transferred to Summit, an HPC machine at the Oak Ridge Leadership Computing Facility (OLCF) in \textit{Workflow 2}, where hyperparameter ranges and datasets are specified. Using Dask, AtomAI, and PyTorch, a parallelized hyperparameter search trains CNN models, with a subset selected based on accuracy and loss metrics. These chosen models are subsequently analyzed and validated in a simpler machine. \textit{Workflow 3}, running on Crusher, a cluster operated at OLCF, evaluates the models by creating evaluation metric plots and calculating additional metrics like mean squared error using the MLFlow data tracking API. Based on the model results, the scientists can deploy them for real-time atom identification or selecting the next measurement point. The experimentation loop concludes when they can interpret the collected data to explain phenomena of interest.

%\ourvspace{}
\subsection{MIDA Implementation}
\label{sec:impl}

In this section, we present our open-source implementation of MIDA~\cite{flowcept} and a detailed description of its development.

\noindent
\textbf{Data modeling.}
MIDA's integrated data view is implemented using a DBMS, employing a task-oriented data design (\cf{section:task_metadata}). Inspired by star schema modeling in traditional data warehouses~\cite{garcia2008database}, we create a data schema focused on task metadata. The \codefont{Task} relation serves as the ``fact table," while other related concepts like environment, system, and user are  dimensions. The \codefont{Task} relation is not normalized, allowing for complex fields such as \codefont{used} and \codefont{generated} to contain inner dictionaries. Additionally, \codefont{start\_telemetry} and \codefont{end\_telemetry} are complex fields capturing resource usage information like GPU memory and IO metrics at the beginning and end of each task. This non-normalized approach, coupled with a flexible schema, can enable faster implementation while maintaining query capability requirements.

% \begin{figure}[!h]
%     \centering
%     \includegraphics[width=0.5\linewidth]{figures/task-schema.pdf}
%     \vspace{-12pt}
%     %\ourvspace{}
%     \caption{%
%         Simplified data schema, focusing on the Task Metadata.
%     }    
%     \label{fig:task_schema}
    
% \end{figure}

\noindent
\textbf{DBMS and MQ System Selection.}
Regarding the user perspective, our prior work~\cite{souza2020workflow} identified diverse analyses that can be done via the integrated data view. These analyses primarily involve querying task metadata and scanning numerous instances with aggregations and filters. These characteristics align with OLAP requirements. 
Regarding the system perspective, support for parallel insertions and updates is crucial, minimizing overhead and ensuring timely availability of data for queries. ACID-compliant task updates are preferred to maintain consistency when multiple observers simultaneously update the same task, such as in Dask observers. These characteristics align with an OLTP workload.
 
In the current implementation, we choose MongoDB for its flexible schemaless features and efficient handling of mixed OLAP and OLTP workloads. MongoDB provides ACID transactions at the single-document level. We model each task instance as a document in the \codefont{Task} collection. Other DBMSs meeting these requirements could be considered.

To implement messaging, the only requirement is that data observers and the data integrator should communicate asynchronously via messages in a common data structure, like a queue. There may be one or many MQ systems and there is no need to use one single MQ system. We opt to use a set of independent and distributed instances of the same MQ system. We use Redis due to its speed, flexibility, and easy deployment, including on supercomputers. Other systems like Kafka or RabbitMQ could be assessed.
% Regarding messaging, MIDA requires asynchronous communication between data observers and the data integrator using a shared data structure like a queue. Multiple MQ systems can be used. In our implementation, we utilize distributed, independent instances of Redis as the chosen MQ system. Redis offers speed, in-memory capabilities, lightweight nature, flexibility, and easy deployment, including on supercomputers. Alternative systems like Kafka or RabbitMQ could also be considered.

\noindent
\textbf{Data Observers' Implementation.}
To develop the data observers, we focus on our motivating use case, which uses popular tools like Dask (a PCS) and MLFlow (a runtime data capture tool).
% We explore our motivating use case to guide the data observers' development, as they vary depending on the use case. With it, we can illustrate the adaptability aspect of our approach, as the use case utilizes popular tools, including a PCS (\ie{} Dask) and a runtime data capture tool (\ie{} MLFlow).
Dask, with its centralized scheduler, only allows the intra-PCS scheduler strategy as it does not provide access to its internal task metadata. 
We extend Dask's plugins to invoke a \codefont{callback} function whenever a task changes state, allowing data observers to send task event messages to the MQ. Since Dask is task-oriented, adapting it to our metadata definition is straightforward, requiring minor adjustments.% such as using an \codefont{enum} domain for task statuses.

The simple example in Listing \ref{listing:dask_code} shows the parallelization of an increment function. Lines 2 and 4 import and register observers using Dask's native plugins. Other than these two lines, there is no change in the parallel code, showing that the observers do not require instrumentation even without the polling strategy. When the map \codefont{incr} executes, the observers track each \codefont{used} input value for the \codefont{n} inputs and each value \codefont{generated} in the output, storing each Dask task, with the corresponding values for \codefont{used}, \codefont{generated}, and others.

% MLFlow uses backend storage for data tracking, so a polling strategy can be applied. MLFlow offers many options for backend storage and this use case utilizes a relational DBMS. Thus, we implement an \codefont{observe} method in the MLFlow adapter, which uses SQL-based polling. The user only needs to provide the routes to access the MLFlow database in a setup file, thus there is no change in the code that uses MLFlow.

MLFlow employs a DBMS for data tracking, enabling the utilization of the polling strategy. This use case uses a relational DBMS. We implemented the \codefont{observe} method in the MLFlow data observer via SQL-based polling. Thus, the user only needs to provide database routes in a setup file without the need to modify the code using MLFlow.
MLFlow uses a \textit{Run} abstraction to store its task metadata, so simple adaptations are needed. Then, whenever MLFlow registers a \textit{Run}, a \codefont{callback} is generated in the data observer that polls the DBMS. The observer adapts MLFlow's \textit{Run} instance into MIDA's task and sends them to the MQ.

\begin{minipage}{0.46\textwidth}
\vspace{0.35em}
\begin{lstlisting}[
    language=python,
    caption={Simple Dask code with observability adapters.},
    label={listing:dask_code},
    keywords={def, return, from, import, print}
]
from dask.distributed import Client
from flowcept import FlowceptDask
client = Client()
client.register_worker_plugin(FlowceptDask())
def incr(n):
  return n+1
futures = client.map(incr, range(1000))
results = client.gather(futures)
\end{lstlisting}  
\ourvspace{}
\end{minipage}

The Dask and MLFlow adapters are designed to be generic, allowing for their reuse in other use cases. They can be further customized to meet specific requirements. Workflow~1 does not use any runtime data capture tool but uses Python's native \textit{logging} library to register the execution evolution in a log file. This log contains the parameter space to control the microscope and the paths of the generated images. Since this is relevant information, we develop a custom data observer for this logging. In the future, if we encounter another use case that already utilizes these popular frameworks, we can leverage their existing adapters, saving valuable time and effort. 

Additionally, the adapters can be customized to support specific \textit{ad-hoc} logging requirements.
Other utilization examples shown as Jupyter Notebooks and a guide to implementing new data observer adapters are available on GitHub~\cite{flowcept}.

% \ourvspace{}
% \subsection{Utilization and Extension}
% \label{section:utilization}

% To utilize the system, one can reuse the existing data observers or implement new adapters by extending and implementing the adapter interface (\cf{sec:data_observer_interface}).
% Then, we employ a single \codefont{configuration.yaml} file with separate keys for each adapter. Details such as routes to a database and polling frequencies can be configured. Before the execution, one needs to start the observer services with the controller API.  Then, without any instrumentation, the observers will work in the background.
% Jupyter Notebooks with examples for Dask and MLFlow data observers are available~\cite{flowcept}.

% stochastic weighted average
% 

 %  Based on the utilization, we will make more query operations available in the API in the future, such as graph traversal queries. 

%\ourvspace{} 
\subsection{Experimentation Setup}	

\begin{figure*}[hbt!]
  \centering
  \begin{subfigure}[b]{0.49\linewidth}
    \includegraphics[width=\linewidth]{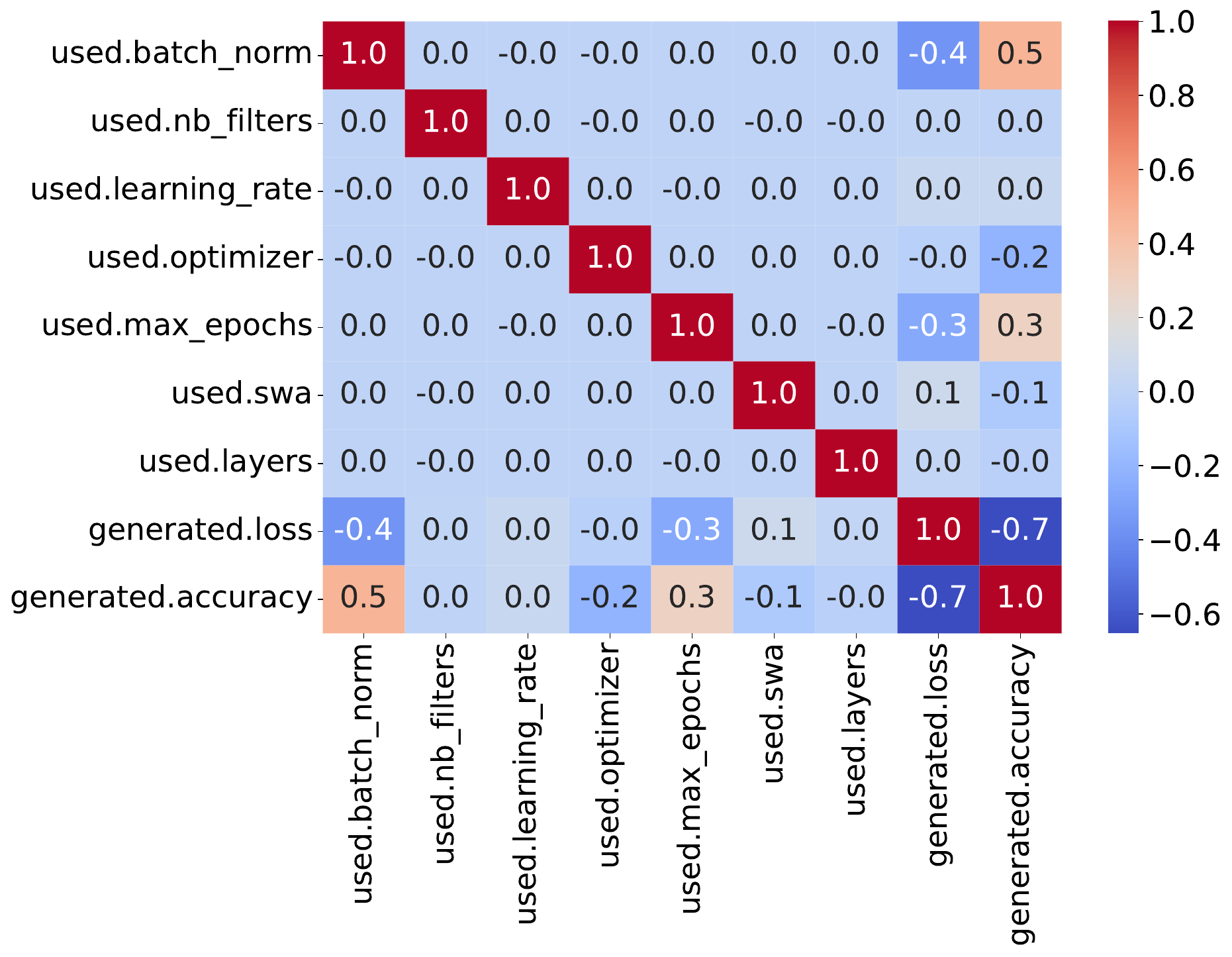}
    \ourvspacefig{}
    \caption{Correlation between Hyperparameters and Model Performance}
    \label{fig:model_perf}
  \end{subfigure}
  \hfill
  \begin{subfigure}[b]{0.49\linewidth}
    \includegraphics[width=\linewidth]{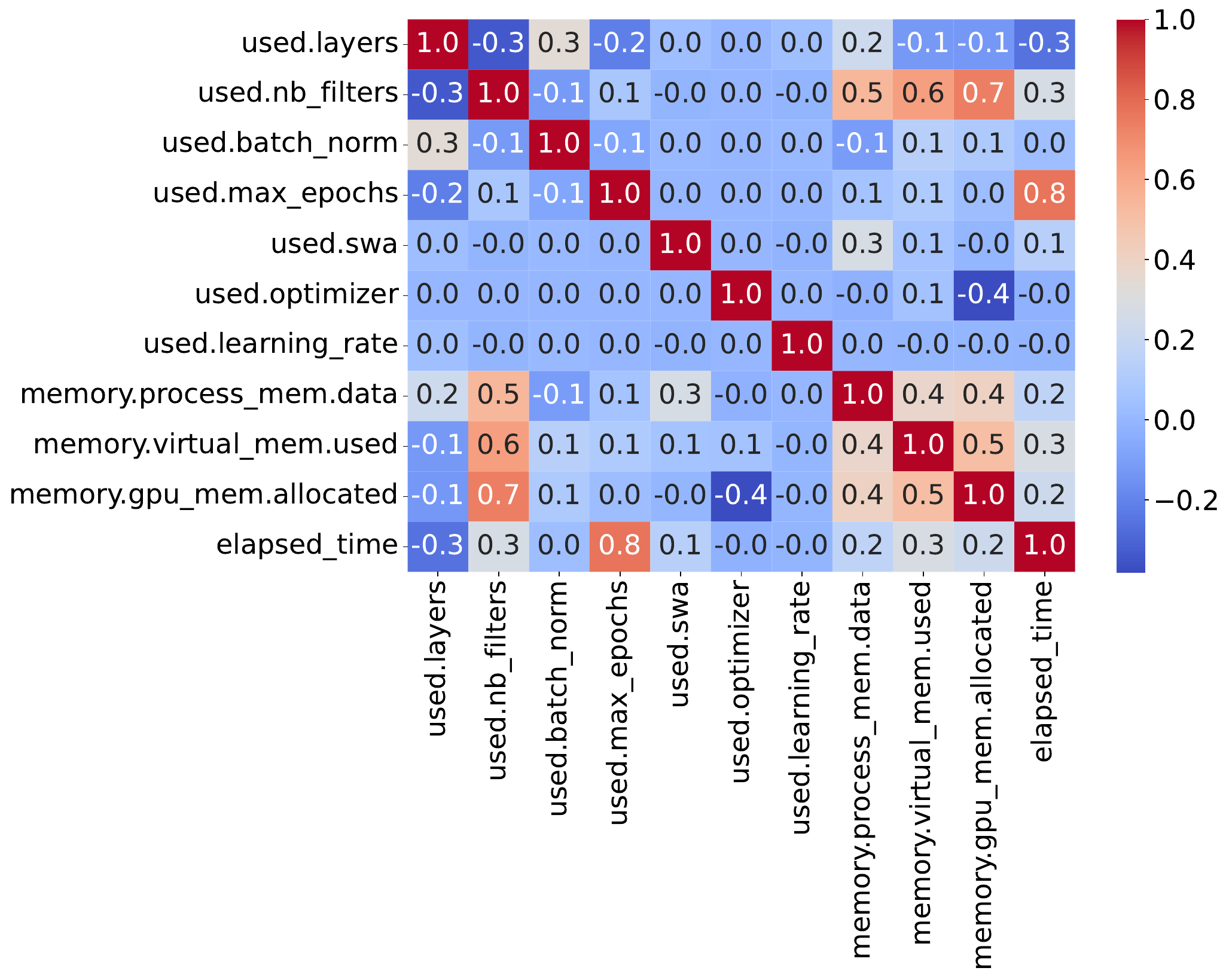}
    \ourvspacefig{}
    \caption{Correlation between Hyperparameters and Telemetry Data.}
    \label{fig:model_telemetry}
  \end{subfigure}
  \vspace{-0.5em}
  \caption{Analyzing correlations between model hyperparameters and KPIs.}
  \label{fig:correlations}
  \aftercaptionvspace{}
\end{figure*}

\noindent
\textbf{Hardware Setup.} The experiments utilize two resources at OLCF: Summit and Crusher. Summit, currently ranked as the 5th fastest supercomputer globally~\cite{top500}, consists of around 4,600 nodes with 2 IBM POWER9 processors and 6 NVIDIA Tesla V100 GPU accelerators per node. Each processor has 22 physical cores, one is reserved for the OS, yielding 42 usable cores per node with 512 GB of DDR4 memory. Crusher comprises 192 nodes, each featuring a single 64-core AMD EPYC 7A53 CPU, 512 GB of DDR4 memory, and 4 AMD MI250X accelerators. In our experimental setup, Summit is primarily used for distributed deep learning computations using Dask in Workflow~2, while Crusher is dedicated to model analysis in Workflow~3. Details on the microscope and the data generated in Workflow~1 are available in previous work~\cite{ziatdinov2020causal}.

\noindent
\textbf{Software Setup.}
Our implementation, v0.1.5, uses MongoDB's container image ibmcom/mongodb-ppc64le:4.4.17 and Redis's 7.0 via Singularity. Workflow~2 uses our data observers for Dask version 2022.12.0, AtomAI v0.7.4, and PyTorch 1.10.2. Workflow~3 uses MLFlow version mlflow-skinny 2.1.1.
%\todo{add Zambeze here for cross-facility data transfers}

\noindent
\textbf{Deployment.}
For Workflow 2, each Summit node runs 6 Dask workers, with one worker per GPU. The Dask scheduler runs on a CPU core in a Summit node. Workflow 2 utilizes 46 nodes, employing a total of 276 GPUs in parallel. MongoDB and Redis operate on Summit's batch node. Workflow 3 runs on a Crusher node, utilizing SQLite for MLFlow. A Jupyter Lab process on Summit's login node acts as an interface for experiment control and data access from Crusher.

%\ourvspace{}
\subsection{Integrated Data Analyses in the Use Case}
\label{sec:integrated_analyses}

We enable three integrated data analyses using our approach. The first analysis involves automatically generating comprehensive documentation of the entire DL model lifecycle, backward from the models generated in Workflow 3 to the parameters used in Workflow 1. The scientist selects the parameter space and retrieves raw microscope images for model training. A Jupyter Notebook calls the Query API, executing queries similar to those in Listing \ref{listing:flowcept_query}. 
Given the input arguments \codefont{campaign\_id} and a number \textit{K} of models, the analysis script generates a document that lists the \textit{K} obtained models with the least mean squared error, which is a metric generated in Workflow 3. For each obtained model, it gets the paths to the  plots generated in Workflow 3 and displays the figures. Then, it navigates in the \textit{used} and \textit{generated} fields into the tasks related to Workflow 2 in the campaign to get the generated accuracy and loss for each of those \textit{K} models along with the used hyperparameters, used datasets, the method chosen to split it into training and validation datasets, collected telemetry metrics (CPU usage and memory usage, GPU memory usage), and elapsed time (\ie{} \codefont{ended\_at} $-$ \codefont{started\_at}), execution environment (\eg{} Summit, Crusher), and user references. Finally, it navigates into Workflow 1, which generated the training and validation datasets used by Workflow 2. Task Metadata in Workflow 1 contains the parameter space and paths to the raw images. 

The second integrated data analysis application further explores the domain data observed in Workflow 2. By calling the Query API, we project all used hyperparameters and generated metrics and plot a correlation matrix in a heatmap. 
By analyzing the correlation between hyperparameters and model performance metrics, users can better understand which choices may deliver more promising results,   create new hypotheses, and try new ideas. Results are in Figure \ref{fig:model_perf}.

\begin{figure*}[hbt!]
    \centering
    \includegraphics[width=1.0\linewidth]{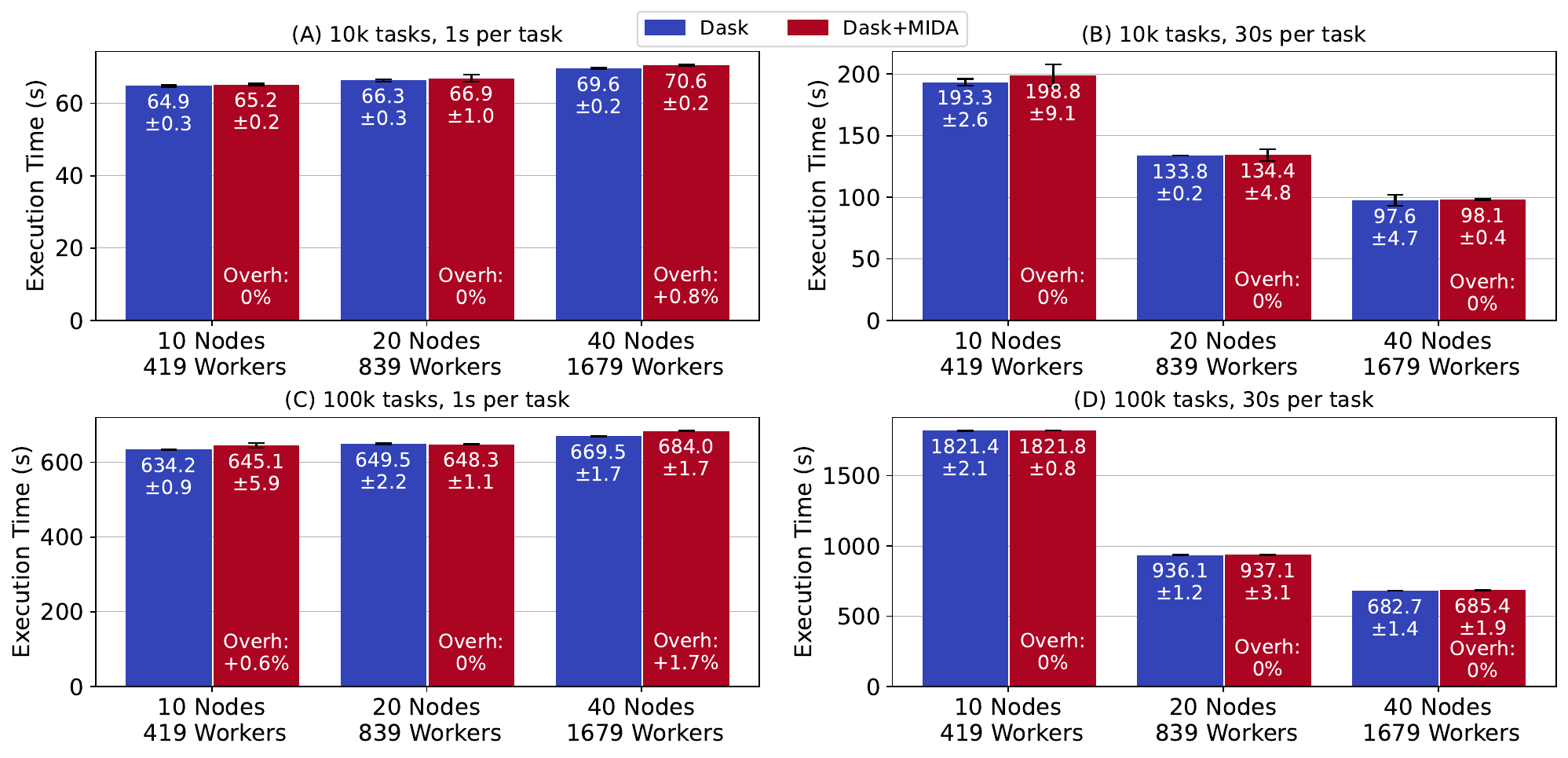}
    \ourvspacefig{}
    \caption{%
        Overhead analysis in Dask workflows with and without data observability and runtime provenance capture.
    }
    \label{fig:overhead}
    \aftercaptionvspace{}  
\end{figure*}

From the results, users can see that  \codefont{batch\_norm} and \codefont{max\_epochs} are positively correlated to the accuracy. As for the \codefont{max\_epochs}, it is expected that the accuracy improves as the training iterates.  Batch normalization, \codefont{batch\_norm}, improves test accuracy by stabilizing the input distributions for each layer during training, providing a regularization effect and accelerating convergence, which enables better learning and generalization from the training to the test data.
% In future work, the users plan to use FlowCept-enabled data analysis while running with a larger variety of hyperparameters and with different sets of microscope images looking for further correlations between hyperparameters and model performance metrics.

This last data analysis combines observed domain data, telemetry data, and elapsed time to steer the workflow. Users aim to identify hyperparameter sets that consume excessive computing resources to avoid execution disruptions caused by memory overflow.
Results are in Figure \ref{fig:model_telemetry}.

From the results, users can see that the hyperparameters \codefont{batch\_norm}, \codefont{nb\_filters}, \codefont{swa}, \codefont{max\_epochs}, and \codefont{layers}  influence both CPU and GPU utilization. However, the \codefont{optimizer} and its \codefont{learning\_rate} have minimal impact on memory consumption. This can be due to additional parameters introduced, while stochastic weight averaging, \codefont{swa}, specifically saves weights at the end of training trajectories. Regarding elapsed time, we confirm that \codefont{max\_epochs} holds the greatest influence. Notably, \codefont{nb\_filters} and \codefont{swa} impact the elapsed time. Interestingly, the number of \codefont{layers} in a neural network shows a negative correlation with elapsed time, possibly due to shallow models underutilizing the GPU.
%Finally, users can also see that CPU and GPU memory utilization are highly correlated.

%\ourvspace{}
\subsection{Overhead Analysis}

Overhead is a major concern for performance-sensitive workflows when the intra-PCS scheduler strategy is applied (\cf{section:main_arch}). We evaluate the data observer overhead in Dask, as it is used in Workflow~2 in the real use case. 
Unfortunately, in this use case, it is impossible to isolate and control variables we need to evaluate, \eg{} ``number of parallel tasks" and ``task duration". Thus, we use a synthetic workflow based on the characteristics of Workflow~2.
After running the \textit{model fit} tasks, which is the most resource-demanding activity, we observe that they last about 30s and the hyperparameter combinations may span over 100,000 tasks. 
Following the performance evaluation practice of other researchers in a similar context~\cite{babuji2019parsl,rosendo2023provlight}, we implement tasks that sleep for a given period.
%, as to measure the overhead at the observers, mainly caused by task state change messages being generated and transmitted over the network, we do not need to mimic heavy computation or GPU utilization. 
We also run a strong scaling evaluation.

To deploy this synthetic workflow, each physical core in a Summit node runs a Dask worker, except for one single core, which runs a Dask scheduler. So, in an execution with 40 nodes, there are 1680 ($=42\times40$) physical cores, 1679 parallel workers (42 workers per node), and 1 centralized Dask scheduler. This deployment allows high task parallelism and creates significant stress both for the task scheduler and for the observers that live both in the PCS's scheduler and in the worker processes, allowing evaluation under high stress.

The workloads are (A) 10,000, 1-s tasks; (B) 10,000, 30-s tasks; (C) 100,000 tasks, 1-s tasks; and (D) 100,000, 30-s tasks. 
We vary from 10 Summit nodes (420 cores) to 20 (840 cores), and 40 nodes (1680 cores), comparing with and without data observability, summing 24 scenarios.  Each workload is a two-map workflow where the number of tasks per map is the total number of tasks divided by two. 
Results are in Figure \ref{fig:overhead}.

We report the execution time medians of a batch of repetitions. For each scenario, for each workload, we repeat at least 5 times and until the 95\% confidence interval of the median is within 10\% of our reported medians. The error bars represent the length of the 95\% confidence interval of the median. We indicate the error bar length below each median time. The indicated overhead percentage is obtained by comparing the extreme values of the interval.
The Dask workflow with and without MIDA is in red and blue, respectively.

We observe that data observability overhead  remains negligible, near 0\%, in all 12 scenarios, even when increasing the number of parallel workers up to 1679. As expected, the largest overhead happens with the largest number of workers, with the largest number of short-lasting tasks. However, the tasks typically last longer than 1 second in reality, such as the \textit{model fit} ones, and given the scale (1680 CPU cores, 100,000 short tasks), we consider that even 1.7\% overhead is still small.
In this scenario, there is more congestion in the network and more pressure in the CPU RAM with 1679 Dask workers generating task event messages in parallel and sending them to Redis. 
To the best of our knowledge, this is the first workflow provenance capture-based system that successfully achieves so low overhead at the scale of a leadership supercomputer. 
Additionally, this is the first experiment with Dask with such a high number of parallel tasks\todo{check Coletti's work, Parsls}. Finally, this is a pioneer work that adds runtime provenance data management to Dask.

Our goal in this experiment is to evaluate our implementation and not Dask's. However, we observe that Dask scales better with 30s workloads, which is consistent with~\cite{babuji2019parsl}.

%\ourvspace{}
\subsection{Lessons Learned and Limitations}

While conducting these experiments, we learned lessons and observed limitations that are worth sharing.

The successful results for maintaining low overhead even in large-scale workloads are due to the system design decisions (\cf{sec:mida}) to address the HPC challenge.
In addition, we found that both MongoDB and Redis were good choices, as both handled dozens of thousands of insertions and updates per second. Our implementation could timely and errorless store the provenance of more than 100,000 tasks of a single Dask workflow, with MongoDB storing more than a million tasks after many experiments. We observed that the index on the \codefont{task\_id} field in \codefont{Task} collection was essential for a successful run since there are multiple tasks' updates at runtime (\eg{} to add the \textit{generated} results). 
Despite Redis being an in-memory system, it checkpoints to disk. We fine-tuned Redis to increase its memory and reduce on-disk accesses.

Moreover, we observed that in-memory buffering event messages and all related downstream operations until the final persistence in the database were essential for efficient performance. For example, while the Dask data observers were generating task state change events, the events are buffered at each observer. When the buffer reaches its limit, the observers use Redis's pipeline insertions to send messages to be received by the data integrator. Then, the data integrator also has an in-memory buffer that groups all insertion and update requests and dispatches a bulk write operation to MongoDB when the buffer reaches its maximum size. 
It is known that although increasing the buffer size can significantly reduce network communication costs, it may reduce the ``at runtime" aspect since the tasks metadata will take longer to be fully persisted in the database.
To address this, in both components, we implemented a dynamic update of the buffer size at runtime based on the frequency of event messages, to further accelerate the ``at runtime" aspect, but we did not assess this in this paper, leaving room for new findings about dynamic buffer sizes and how they affect the ``at-runtime" aspect in future work.

Altogether, these design decisions and fine-tunes in Redis collaborated to a successful run with low overhead. However, we did not assess which specific design decision or fine-tuning impacted the overhead management the most. More experiments isolating the choices and fine-tunes could provide more insights to future system researchers and practitioners.

We also found that MongoDB and Redis were good choices in terms of development speed and deployment. Their schemaless feature provided flexibility to keep a fast pace of development maintaining the runtime steering capabilities. 
Although deploying DBMSs on HPC machines may be complex,
deploying is straightforward thanks to containerization technology (\eg{} Singularity), which is often available in those machines.

Finally, focusing on Task metadata as the finest element to manage enabled high capillarity to capture the dataflows, allowing both fine-grained depth-first data analysis in single-workflows and coarse-grained breadth-first analysis in multi-workflows, as shown in the analyses in Section \ref{sec:integrated_analyses}.
% \ourvspace{}
\section{Related Work}

\noindent
\textbf{Provenance in Parallel Computing Systems.}
Workflow management systems are known for having incorporated provenance data management capabilities for years. For example, 
systems like Swift/T~\cite{wozniak2013swift}, FireWorks~\cite{jain2015fireworks}, and Pegasus~\cite{deelman2021pegasus} collect and store provenance data in log files or DBMSs.
However, they aim at managing single-workflow executions, thus they alone cannot provide for multi-workflow analyses. Other PCSs, like big data processing systems such as Spark, have extensions for provenance management~\cite{guedes2020capturing} but are also made for single-workflows. 
Many other PCSs do not have any provenance support, like Dask. Our approach is made to integrate with systems that either already manage provenance or not. In the case of systems that do not manage provenance, the first step is to implement a data observer, which adds the basic functionality of single-workflow runtime provenance management. Then, once with single-workflow provenance capabilities, these systems can be integrated into the MIDA system to enable multi-workflow analysis.

\noindent
\textbf{Runtime Data Capture Tools.}
There are several runtime data capture tools available.
We can categorize them by whether they require instrumentation or not and whether they are domain-agnostic or specific.
For example, DfAnalyzer~\cite{dfanalyzer_2020} and YesWorkflow~\cite{mcphillips2015yesworkflow} are domain-agnostic and require instrumentation. Other tools, like NoWorkflow~\cite{pimentel_noworkflow_2017}, are domain-agnostic and do not require instrumentation. 
Tools like PROV-IO~\cite{prov_io} are flexible to both automatically capture data from specific operations, such as IO calls, and allow custom instrumentation.
MLFlow captures ML-specific provenance and requires instrumentation, whereas other works~\cite{cunha_2021_context,mci_Kerzel2023} capture data science-specific provenance from Notebooks without instrumentation. There are tools for runtime telemetry data collection by instrumenting scripts~\cite{carns2011understanding}.
These tools are also made for single-workflows, meaning they are not designed to provide end-to-end analyses spanning across various workflows in a multi-workflow. They can be complementary to each other by integrating into our MIDA approach.

\noindent
\textbf{Multi-workflow Systems.}
Some systems provide for multi-workflow orchestration capabilities~\cite{simitsis2012optimizing,rogers2013bundle}, but they do not aim at integrating data from various data sources at runtime, and thus do not provide an integrated data view.
Our previous work, ProvLake \cite{provlake_escience_2019}, can provide this integrated data view at runtime for multi-workflows. Nevertheless, lacking an adapter-based design, it is not designed to integrate with provenance data from various PCSs or runtime data capture tools, failing to address the Multitude of Tools challenge. Without data observability capabilities, it also requires the instrumentation of all scripts, which may be cumbersome to adopt in certain use cases.
Despite this, as the other runtime data capture tools, ProvLake can be integrated into the MIDA system, thus not competing but extending MIDA's capabilities.

\noindent
\textbf{Data Observability Tools.}
Systems like DataHub~\cite{datahub} and the work \cite{observability_sigmod2021} show how data observability  benefits monitoring and user steering, enabling humans in the loop of complex large-scale distributed system deployments. Nevertheless, these solutions are not workflow-oriented, not even for single-workflows. This prevents performing end-to-end data analyses, such as the ones we show in our experimental validation section.

Other solutions aim at integrating domain-agnostic provenance databases with a standardized schema~\cite{missier2010linking,hardy2022sciproflow}, but they do it after the fact. MIDA is domain-agnostic, capable of representing various domains, and is designed to provide efficient data integration at runtime.

%\ourvspace{}
\section{Conclusion}

This paper introduces an innovative approach for runtime multi-workflow integrated data analysis, MIDA, building on provenance techniques, data observability, and an adapter-based system design. The primary objective of MIDA is to enable seamless data integration coming from various tools, including both parallel computing systems, with and without existing provenance support, and runtime data capture tools, regardless if they are provenance-specialized tools or not. To facilitate cross-facility scientific discovery, we proposed a generic system architecture for data integration. By conducting experiments for deep learning in a real materials science use case, we demonstrate the effectiveness of the approach. 

The multi-workflow provenance traces  benefit users by helping them have a better end-to-end understanding of their experiments through user steering, enabling combined analyses from the trained models until the parameter space the scientist chose to generate raw data, across three workflows. MIDA enabled higher transparency of the generated AI models and better reproducibility by recording the provenance of the results, contributing to a more responsible AI development. 
Finally, the implementation exhibits near-zero overhead even with HPC workloads comprising up to 100,000 parallel tasks on 1680 cores, as evidenced through experiments using Dask, a system that lacked provenance support before this work.

% Future work stuff:
% - In this paper we focused on building the infrastructure to enable the integrated data analysis. For future work we plan to expand the analytical capabilities by adding:
% - raw data integration, more specific ML model semantics and evaluation, graph querying capabilities...
% - Native integration with various profiling tools
% Use WfCommons for the task metadata and WfBench for the synthetic exps
% - better integration with SteeringActions (GUI activity)
% Future work: investigate other more complex provenance data models

\medskip
{\scriptsize
\noindent \textbf{Acknowledgements.}
To Ketan Maheshwari, Joshua Brown, and Mark Coletti (ORNL) for their help. 
This research used resources of the Oak Ridge Leadership Computing Facility 
at the Oak Ridge National Laboratory, which is supported by the Office of 
Science of the U.S. Department of Energy under Contract No. DE-AC05-00OR22725.
Support for DOI 10.13139/ORNLNCCS/1773704 dataset is provided by the U.S. Department of Energy, project Sm-BFO STEM under Contract DE-AC05-00OR22725. Project Sm-BFO STEM used resources of the OLCF at Oak Ridge National Laboratory, supported by the Office of Science of the U.S. Department of Energy under Contract No. DE-AC05-00OR22725.
}

%% References
\bibliographystyle{IEEEtran}
%\nocite{*}
\bibliography{references}
\end{document}